\documentstyle[12pt,pra,aps,epsf]{revtex}
\tighten
\begin{document}
\draft
\title{Entangling power of quantized chaotic systems}
\author{Arul Lakshminarayan}
\address{Physical Research Laboratory,\\
Navrangpura, Ahmedabad,  380 009, India.}   
\maketitle
\newcommand{\newc}{\newcommand}
\newc{\beq}{\begin{equation}}
\newc{\eeq}{\end{equation}}
\newc{\kt}{\rangle}
\newc{\br}{\langle}
\newc{\beqa}{\begin{eqnarray}}
\newc{\eeqa}{\end{eqnarray}}
\newc{\longra}{\longrightarrow}

\begin{abstract}

We study the quantum entanglement caused by unitary operators that
have classical limits that can range from the near integrable to the
completely chaotic. Entanglement in the eigenstates and time-evolving
arbitrary states is studied through the von Neumann entropy of the
reduced density matrices. We demonstrate that classical chaos can lead
to substantially enhanced entanglement. Conversely,
entanglement provides a novel and useful characterization of quantum
states in higher dimensional chaotic or complex systems.  Information about
eigenfunction localization is stored in a graded manner in the Schmidt
vectors, and the principal Schmidt vectors can be scarred by the
projections of classical periodic orbits onto subspaces.  The
eigenvalues of the reduced density matrices are sensitive to the
degree of wavefunction localization, and are roughly exponentially
arranged. We also point out the analogy with decoherence, as 
reduced density matrices corresponding to subsystems of fully 
chaotic systems are diagonally dominant.

\end{abstract}

\section{Introduction}

Entanglement has been studied since the early days of quantum
mechanics and provides the essential ingredient of EPR phenomena \cite{Peres}.
It results when the state of a system, which is composed of
at least two subsystems, cannot be written as a product of states that
reside entirely in the subsystems. This leads to the well known unique
quantum correlations that exist even in spatially well separated pairs
of particles. More recently entanglement has been discussed
extensively in the context of quantum information theory. It has been
recognized as a quantum resource for quantum superdense coding and
quantum teleportation, while helping make quantum computations
qualitatively superior to classical ones \cite{Steane}. 

Usually the entangled subsytems are distinct identical particles and
the entanglement is created by symmetrization, such as in the spin
singlet state of a fermionic pair. However, entanglement may of course
be created during time evolution due to conventional interactions of a
potential nature. In this case, the ``subsystems'' may also be the
many degrees of freedom of the same particle. This is more general
than the special case wherein the interaction preserves the
permutation symmetry between the say $d$ (formal) degrees of freedom
and may therefore be interpreted as representing $d$ identical
particles in one dimension.

Presented with the inexpensive resource of unentangled states, unitary
operations may be devised to create potentially useful entangled
states. Thus an understanding of the range of entanglement produced by
the wide range of unitary operations is of interest.  If these unitary
operators are generated from Hamiltonians, then a natural question
that arises is the connection between the dynamics generated by the
Hamiltonian and the entanglement produced. In particular it has been
recognized for some time now that the two extreme cases of classical
dynamics, namely the completely integrable and the completely 
chaotic \cite{LL}, leave remarkably different traces on quantization 
\cite{QCbooks}.  
In this paper we enquire into the possible effects of quantum chaos on
quantum entanglement. A related issue is that of decoherence
\cite{Joos} and our treatment is equivalent to treating the
environment as a subsystem of a quantum chaotic system.

Certain finite unitary matrices have been studied extensively in the
quantum chaos literature and represent a rich source of operators with
a wide range of well understood dynamical behaviors in the classical
limit .They may for instance arise naturally while quantizing
symplectic maps on the torus \cite{HB,BV,Izrailev}. However, most of
the studies have been limited to the two-torus, with essentially
one-degree of freedom, while entanglement studies necessarily deals
with the entanglement between two subsystems. Thus, while two-degree
of freedom Hamiltonian flows lead to one-degree of freedom maps, we
need to consider at least a four-dimensional symplectic map to study
entanglement directly.

We use a coupled quantum standard map as the model system. The
standard map is well understood classically, there is a smooth
transition from the regular to the chaotic, and has been the object of
several quantum investigations as well (reviewed in \cite{Izrailev}), 
including an experimental 
realization \cite{Raizen}.  The
coupled map has been studied classically, principally, to understand
Arnol'd diffusion
\cite{4DClass}. Thus our results below also represent new results in
the largely sparse literature on higher dimensional ($d>1$ for maps,
$d>2$ for flows) quantum chaotic systems. In particular we explore the
usefulness of entanglement as a measure of quantum chaos for such
systems.  Our results, largely exploratory and numerical in nature,
quantify what is intuitively (and certainly linguistically) evident:
the connection between entanglement and chaos. The larger the chaos,
more the entanglement that is induced by the dynamics, as opposed to
entanglement due to symmetries. However we also find interestingly
that the reduced density matrices store information about localization
of states in a systematic manner.

Consider a bipartite quantum system whose state space is ${\cal
H}\,=\,{\cal H}_1 \, \otimes \, {\cal H}_2$ where $\mbox{dim} \, {\cal
H}_i\,\, = N, \; (i=1,2)$. The unitary operator acts on vectors in the
space ${\cal H}$ where $\mbox{dim} \, {\cal H}\, =\, N^2$.  We consider the
case where the two subsystems have the same dimensionality.  Let $|\psi
\kt
\, \epsilon \, {\cal H}$, and the reduced density matrices be $\rho_1=
\mbox{tr}_2(|\psi \kt \br \psi|)$ and $\rho_2= \mbox{tr}_ 1(|\psi \kt
\br \psi|)$, where the first matrix is obtained by tracing out the second
degree of freedom and the second by tracing out the first.  The von
Neumann entropy $S$, referred to in the rest of the paper as simply
entropy, of the reduced density matrices measures the entanglement of
the pure state $|\psi \kt$ in an essentially unique manner.
\beq
S\, =\, -\mbox{tr}_1(\rho_1\, \log(\rho_1))\,=\,  -\mbox{tr}_2(\rho_2\, 
\log(\rho_2)).  
\eeq
The close analogy between this entropy and thermodynamic entropy has
been noted and discussed earlier \cite{PopRoh97}.

The eigenvalues of the reduced density matrices form the 
(square of the) coefficients of the Schmidt decomposition.
This decomposition expresses the state in the full Hilbert space as
a linear combination of $N$ product states instead of the evident $N^2$.
If we were to write the eigenvector for instance in the position basis 
and in the Schmidt decomposed forms we would have
\beq
|\psi \kt \, =\, \sum_{n_1,n_2} c_{n_1n_2}\, |n_1\kt |n_2 \kt \, =\, 
\sum_{j=1}^{N} \sqrt{\lambda_j}\, |\phi_j \kt^{S} |\phi^\prime _j \kt ^{S}.
\label{schmidt}
\eeq
A simple proof of this is found in the Appendix of \cite{Albrecht}. 
Thus there is a basis in each Hilbert space such that each basis
vector in ${\cal H}_1$ is uniquely correlated with one in ${\cal H}_2$
as far as the state $|\psi \kt $ is concerned. The states $|\phi_j
\kt^{S}$ and $|\phi^{\prime}_j \kt^{S}$ are the eigenvectors of the
reduced density matrices, and form the basis of the Schmidt
decomposition. The Schmidt decomposition is not merely a mathematical 
convenience but is understood to provide a deeper understanding 
of correlations between subsystems.

In the literature on quantum chaos or localization \cite{Izrailev,SSL}
, the Shannon entropy
has often been calculated, and for the above state it is given by:
\beq
S_{shan}(|\psi \kt)\,=\,-\sum_{n_1,n_2} \, |c_{n_1 n_2}|^2 \,
\log( |c_{n_1 n_2}|^2).
\eeq
This quantity is of course basis dependent and vanishes if the basis
is chosen to have a direction along $|\psi \kt$. The von Neumann
entropy or entanglement is also basis dependent, but is to an
important extent immune to arbitrariness by being invariant under {\it
local} unitary transformations, even such not being the case for the
Shannon entropy.  Local transformations are those that act only on
individual subspaces.

\section{Four Dimensional Standard Map}

We now define the four-dimensional standard map \cite{4DClass}. It is
composed of two pendulums that are periodically kicked and are also
coupled to each other.  The symplectic transformation of the phase
space variables $(q_1,q_2,p_1,p_2)$, connecting states just before two
consecutive kicks, separated by unit time, is the classical map:
\begin{mathletters}
\beqa
q_1^{\prime}&=&q_1\,+\,p_1^{\prime}\\
p_1^{\prime}&=&p_1\,+\,\frac{K_1}{2\pi}\, \sin(2 \pi q_1)\,+\,
\frac{b}{2 \pi}\, \sin(2 \pi(q_1+q_2))\\
q_2^{\prime}&=&q_2\,+\,p_2^{\prime}\\
p_2^{\prime}&=&p_2\,+\,\frac{K_2}{2\pi}\, \sin(2 \pi q_2)\,+\,
\frac{b}{2 \pi}\, \sin(2 \pi(q_1+q_2)).
\eeqa
\label{clmap}
\end{mathletters}

The phase space is restricted to the unit four-torus $T^4$, and
therefore $\mbox{mod}\, 1$ operations are understood in all of
Eqs.~(\ref{clmap}).  If $b=0$, the system falls into two uncoupled
standard maps. In this limit much is known of the dynamics \cite{LL}; briefly,
if $K\,=\,0$ (referring now to either $K_1$ or $K_2$), the dynamics is
integrable, while at $K \approx 1$ the last KAM rotational tori
breaks, heralding large scale diffusion in phase space, for $K<5$, the
phase space is that of a typical Hamiltonian system, a mixed phase
space with both regular and chaotic regions.  When $K>>5$, the
dynamics is practically completely chaotic with possible appearances
of very tiny stable islands through tangent bifurcations.  For $b \ne
0$, little is known, due to the dimensionality of the phase space,
although this map has been used in studies of Arnol'd diffusion. We
suggest, and substantiate below, that in cases such as these where
finite unitary matrices may be constructed as quantization, the
quantum maps can be used to actually find transitions to classical
chaos. As we noted for the general case earlier, in the case $K_1\,=\, K_2$,
the system possesses permutation symmetry between the two degrees
of freedom and may be interpreted as {\it two} interacting particles
in a one-dimensional standard map external potential.

The quantization of the symplectic transformation in Eq.~(\ref{clmap})
is a finite unitary matrix, whose dimensionality is $N^2$, and
$N=1/h$, where $h$ is a scaled Planck constant. The classical limit is
the large $N$ limit.  The quantization is straightforward as there
exists a kicked Hamiltonian generating the classical map.  The quantum
standard map on the two-torus in the position representation is
\beqa
\label{2dqmap}
U(n^{\prime},n; K_1,\alpha,\beta)\,=\, \frac{1}{N}\,  \exp\left(- i N
\frac{K_1}{2 \pi} \cos(\frac{2\pi}{N}(n+\alpha))\right) \nonumber \\ \times
\, \sum_{m=0}^{N-1}\exp\left(-\frac{\pi i }{N} (m+\beta)^2\right) \, 
\exp\left(\frac{2 \pi i}{N} (m+\beta)(n-n^{\prime})\right).
\eeqa
The position kets are labeled by $n\,=\, 0,N-1$ and the position
eigenvalues are $(n+\alpha)/N$ while the momentum eigenvalues are
$(m+\beta)/N$, $m=0,\ldots,N-1$. Here $\alpha$ and $\beta$ are real numbers in
$[0,1)$ which represent quantum boundary conditions and are convenient
devices for breaking phase space reflection symmetry (the phase $\alpha$)
and time reversal symmetry (the phase $\beta$).  The four-dimensional
quantum map is but a simple extension:
\beqa
\label{4dqmap}
\br n_1^{\prime} n_2^{\prime}|{\cal U}|n_1 n_2 \kt \,=\, 
U(n_1^{\prime},n_1; K_1,\alpha,\beta)\, 
U(n_2^{\prime},n_2; K_2,\alpha,\beta)\,\nonumber \\
\times \exp\left( - i N  \frac{b}{2 \pi} \cos(\frac{2 \pi}{N}
(n_1+n_2+2 \alpha))\right).
\eeqa
${\cal U}$ is a unitary matrix in ${\cal H}$, and will induce  mixing 
between the two subsystems. We have assumed the quantum phases in both the
subsystems to be identical. Throughout this paper we use $\alpha=0.35,
\, \beta=0$ as the quantum phases. 

\section{Results}
\subsection{Stationary State Properties}

If $b \neq 0$ an unentangled initial state, such as $|n_1,n_2 \kt \,
\equiv \, |n_1\kt \otimes |n_2 \kt$, would eventually get entangled by
the repeated action of the unitary operator ${\cal U}$. The entangling
properties depend on the entanglement already inherent in the
stationary states or eigenstates ${ |\psi_i \kt, \, i=1,\ldots,N^2 }$,
of ${\cal U}$.  Thus we first calculate the average entropy of the
eigenstates when $K_1 \, =\, 0.1,\, K_2\,=\, 0.15$ as a function of
the coupling constant $b$.  At these values of $K_i$ the uncoupled
standard maps are almost wholly regular. We calculate
\beq
\label{avent}
{\overline S}\,=\, \frac{-1}{N^2} \sum_{i=1}^{N^2} \, \mbox{tr}_1(\rho_{1i} 
\log(\rho_{1i})), \;\; \rho_{1i}\,=\, 
\mbox{tr}_2 (|\psi_i \kt \br \psi_i |)
\eeq
where
$\rho_{1i}$ is the reduced density matrix after tracing out the 
second degree of freedom. The quantity ${\overline S}$ is a gross
quantity averaged over the entire spectrum, and gives an idea of the 
average entanglement we can expect on using the operator ${\cal U}$.
In Fig.~(\ref{figavent}) we see the entropy increasing from zero at $b=0$ 
and attaining a nearly constant value beyond $b \approx 3$.

The increase in the entropy proceeds along with a gradual increase of
chaos in the system, flattening out after considerably uniform chaos
has been achieved, a fact that is confirmed by iterating the classical
map Eq.~(\ref{clmap}).  In fact this suggests a compact way of
exploring the {\em classical} transition to chaos which is otherwise
mired in problems of visualizing four-dimensional sections. Thus
entanglement is clearly a function of the nature of the underlying
classical dynamics. That the entropy increases as an approximate power
law before flattening out is shown in the inset. Roughly we get
${\overline S} \,\approx \, b^{.4}$.  We note here that the Shannon
entropy would behave differently, as even when $b=0$ there is a
non-zero Shannon entropy in general. If there is large scale chaos in
the subsystems (as in our model if $K_1$ and $K_2$ are greater than
five) it will be reflected as a large Shannon entropy; however the
entanglement would be zero. For large coupling between the
subsystems, both the entropies appear to be well correlated.

Subject to the constraint that $\mbox{tr}_1(\rho_{1i})\,=\, 1$, the
maximum entropy is $\log(N)$, and corresponds to the ``microcanonical
ensemble'' with all the eigenvalues $(\lambda_j, \, j=1,\ldots,N)$ of
$\rho_{1i}$ being equal to $1/N$.  The entanglement entropy induced by
the dynamics of quantum chaos falls short of this and in fact the
eigenvalues if arranged in decreasing order are exponential and
reminiscent of the ``canonical'' ensemble; see discussion below.  From
the data of Fig.~(\ref{figavent}) it appears that at saturation
${\overline S} \, \approx \,
\log (0.59 N)$. Thus roughly $0.59 N$ pairs of correlated states from
the two subspaces make up a typical state. The meaning of this is that
of the optimal minimum number of components present in the full state
if we are given the freedom to choose a basis from each subspace.  If
we are given such a choice in the full Hilbert space we would have
just one component with the basis having one of its directions aligned
along the eigenstate, and the von Neumann entropy of the pure state is
zero.  This may also be to compared to a $M$ dimensional random matrix
eigenvector, belonging to the Gaussian Orthogonal ensemble (GOE) 
\cite{PandeyRMP} whose Shannon entropy is approximately $\log (0.5 M)$. 

We turn now to a somewhat more detailed study of the entanglement
inherent in individual eigenstates. In Fig.~(\ref{figent}) is plotted
the individual entropies corresponding to all the states of a fairly
chaotic system.  While most of the states have already achieved the
entropy corresponding to the saturation value of $\log(.59 N)$ (small
dots), there are many states that  are prominently low in
entanglement (those with an entropy less than $3$ is marked with a 
circled dot).  The Shannon entropy, not displayed here, of the (small
dot) states is to a large accuracy $ \log(0.5 N^2)$, (as the
dimensionality is $M=N^2$), while there are also minima that largely
match with those in Fig.~(\ref{figent}) and therefore these are
expected to be localized states.  In Fig.~(\ref{figefn}) is shown two
wavefunctions, one a typical chaotic state and the other a localized
state corresponding to the first prominent minima of the entanglement
entropy shown in Fig.~(\ref{figent}) ($i=212)$. 

Many of the low entropic states are similar to the localized state in
Fig.~(\ref{figefn}), and suggests a ``scarring'' \cite{Heller}
mechanism. In fact this results from a scarring due to a fixed point
of {\it mixed} stability, and may be called semi-scarred if one is to
strictly define scarring as due to unstable, hyperbolic orbits
\cite{Kaplan}. The initial condition $(q_1=0.5,p_1=0,q_2=0,p_2=0)$ is
a fixed point for all values of the parameters. When $b=2$ the
eigenvalues of the Jacobian at this point has a real pair,
corresponding to hyperbolic motions and a complex conjugate pair
corresponding to stable motions. This fixed point is poised to
completely lose stability just after 2 (at $\approx 2.01)$.  While
providing a new example of scarring in higher dimensional systems due
to orbits of mixed stability type, this shows that entanglement is
sensitive to eigenfunction localization or scarring.  We might expect
this to be true for a large class of strongly scarred states.

The structure of the reduced density matrices corresponding to these
two states are displayed in Fig.~(\ref{figrdm}) and reveal the
connection to decoherence phenomena.  The reduced density matrices are
typically diagonally dominant in the presence of large scale chaos.
It is interesting that the localized states are more ``cleanly''
diagonal than the typically delocalized state. There is a rather rapid
transition from a non-diagonal to a predominantly diagonal density
matrix as the system undergoes a transition to chaos; even mixed phase
spaces seem to lead to sufficiently diagonal density matrices. This is
shown in Fig.~(\ref{figavRDM}) where the average (of the absolute
value squared) of the density matrices corresponding to the entire
spectrum is shown.  The reduced density matrices of non-stationary
states also tend to a diagonal structure very rapidly when the system
is chaotic, as will be demonstrated later.

Due to the reduced density matrices being diagonally dominant, it is
not obvious if they themselves have random matrix properties.  However
the state in the full Hilbert space possesses properties of
eigenvectors of random matrices of the GOE (or COE) type. We may then
use this to understand the typical diagonal nature of the reduced
density matrices. The diagonal term is (dropping the state index $i$ in 
favor of the indices indicating the position in the reduced density matrix):
\beq
\rho_{1nn}\,=\, \sum_{k}\, |\br n\, k|\psi \kt |^2.
\eeq
Typical (ensemble averaged) value of $|\br n \,k|\psi \kt |^2$ is $1/N^2$,
which one can see easily from merely the normalization condition. Therefore
the typical diagonal element is of order $1/N$. The ``strength'' of the
diagonal elements as measured in the spectral average, and shown in 
Fig.~(\ref{figavRDM}) will be 
\beq <|\rho_{1nn}|^2> \, \sim \, \frac{1}{N^2} \eeq
The off-diagonal element is given by
\beq
\rho_{1n n^{\prime}}\,=\,  
\sum_{k}\, \br n \,k|\psi \kt \br \psi|n^{\prime} \,k  \kt. 
\eeq
The ensemble average of this quantity vanishes. The strength of these elements
is 
\beq 
<|\rho_{1n n^{\prime}}|^2 >\,\sim\,<\sum_{k}\, |\br n\, k|\psi \kt 
\br \psi|n^{\prime} \, k  \kt|^2>\, \sim\, N \, \frac{1}{(N^2)^2}\,
=\, \frac{1}{N^3},
\eeq 
where we have used the GOE result \cite{PandeyRMP} that 
\beq
<<|\br i|\psi \kt 
\br \psi|i^{\prime}  \kt|^2>> \, =\, \frac{1}{M (M+2)}\, 
\approx\,\frac{1}{M^2}; \;\; (i\ne i^{\prime}),
\eeq
and $M=N^2$ is the dimensionality of the matrix.
Thus the average off-diagonal element will be smaller than the
diagonal by a factor of $\sqrt{N}$. This is borne out to a large
extent by the numerical results.

Is there any advantage in using the reduced density matrix, rather
than say the Shannon entropy as a measure of eigenfunction properties,
in particular of localization? Clearly the reduced density matrix
contains much more by way of information, and the von Neumann entropy
derived from it is just one piece of information that gives a global
idea of localization. If we have the complementary reduced density
matrix obtained by tracing out the first degree of freedom we will
have complete information about the wavefunction, via the Schmidt
decomposition. We now study the spectral properties of the reduced
density matrices and observe how information about localization is
stored in a remarkably graded manner.

The eigenvalues $(\lambda_j; j=1, \ldots, N )$ of the reduced density
matrix, assumed to be arranged in increasing order, also naturally
contain information about localization.  As we noted earlier these
eigenvalues fall off exponentially as seen in Fig.~(\ref{figeval}).
The localized state seems to have at least two exponential scales while
the generic delocalized state has only one. For the latter class of
states we find numerically that
\beq
\lambda_j \, \sim \, \mbox{const.}\, \exp(-\frac{\gamma j}{N}),
\eeq
where $\gamma$ is an $N$ independent constant. While this is valid for
the most significant few of the eigenvalues, there is a clear
deviation of this for smaller eigenvalues and the exponential law
seems modulated by a polynomial one that we will not investigate in
more detail.

The role of the ``temperature'' is played by the Hilbert space
dimensionality $N$.  This may be compared to the ``temperature of the
eigenvalue gas'', which measures the equilibrium (fully chaotic)
distribution of the level velocities, which is also proportional to
$N$, but the proportionality constant (the ``Boltzmann constant'') may
be a measure of system specific classical correlations \cite{LCS}.
The localized state seems to be cleanly split into two parts, one with
a localization dominant part and the other which behaves like a
generic extended state. The degree to which there are two
distinguishable scales depends on the particular localized state. Thus
the information about the state's localization is present in the first
few Schmidt states of the reduced density matrix.

We now study the corresponding eigenvectors of the reduced density
matrices.  These are the Schmidt states $|\phi_j \kt^{S}$ in the
expansion of Eq.~(\ref{schmidt}) for eigenstates. Since these are in
$N$ dimensional subspaces spanned by either degree of freedom, they
correspond classically to ``projections in the $(q_i,p_i)$ space''.
Here $i=1,2$ while tracing out the second or the first degree of
freedom respectively.  We see them in two ways: one is the
usual position basis, the other is a phase-space representation, such
as a Husimi distribution. The latter will reveal phase-space scarring
effects, provided of course that we know enough about the classical
dynamics. 

The position basis representation of some of the principal
eigenvectors or the Schmidt vectors of the reduced density matrices of
two states is shown in Fig.~(\ref{figvecrdm}).One of these states is a
typical non-localized state while the other is the localized state
discussed above.  While the Schmidt vectors of the non-localized state
are essentially unremarkable and appear to be random, those that
belong to the localized state are themselves localized and appear
arranged in the manner typical of eigenstates with increasing node
numbers.  Beyond the six states shown here, Schmidt vectors of the
localized state also look random and indistinguishable from the
delocalized state, as is already evidenced in the sixth state. This
agrees with the fact that the eigenvalues of the Schmidt vectors too
seem to share common trends beyond this point. Thus information about
state localization is stored in the Schmidt vectors in a graded
manner.

The Husimi of the Schmidt vectors will reveal more about the classical
structures that influence the localization. We have asserted
previously that the localized state we have been analyzing is
influence by the fixed point $(0.5,0,0,0)$ and while this is plausible
from the state vector in the position basis already displayed, it is
confirmed by the Husimi of the Schmidt vectors.
Thus we plot 
\[ W(q,p)\,=\, |\br q,p|\phi\kt^{S}|^2
\]
in Fig.~(\ref{fighusm}) where $|q,p \kt$ is a coherent state on the
two-torus as developed in \cite{Saraceno}. We have also plotted the
complementary Schmidt vectors from tracing out the first degree of
freedom and will therefore give the $(q_2,p_2)$ coordinates of any
classical structure.  The fixed point $(0.5,0,0,0)$ is clearly seen
projected onto the two subspaces in the Husimis, thus confirming our
earlier statement that the state is ``scarred'' by this orbit. In general
we may expect that periodic orbits scarring the states will be seen in their
projections in the Husimis of their principal Schmidt vectors. 

The full Husimi distribution is of course four-dimensional and taxes
our visualization abilities. The above may be compared to a similar approach
that has already been in use when four dimensional Husimis of
two-dimensional eigenfunctions of chaotic oscillators with two-degrees
of freedom were analyzed via their ``quantum surface of sections''
\cite{Santh}. The difference is that the Schmidt states provide a much
more complete and systematic way of analyzing higher dimensional
wavefunctions than the somewhat ad-hoc constructions thus far in use.
It has been shown for two-dimensional maps and two-degree of freedom
flows (which are equivalent) that the zeros of the Husimi
distribution, dubbed as ``stellar representations'', provide a unique
description of the eigenfunctions \cite{LebVoros}. The stellar
representation of Schmidt vectors may provide a way of avoiding
complex functions in two variables; as each vector and its correlated
partner will have $N$ zeros each and there are $N$ such pairs, we
would have a total of $2 \, N^2$ zeros. Of course these in themselves
are not sufficient to specify the state (as we need the eigenvalues of
the reduced matrix as well), but we have seen that important information
about the states is already present in the Schmidt vectors and must be
reflected in their zeros. Calulations not presented here indeed confirm 
this.

\subsection{Time Dependent Properties}

We dwell briefly on time dependent properties.  While for
wavefunctions on the full Hilbert space, time dependent properties may
largely be derived from the stationary one, the situation is not
entirely obvious when we restrict ourselves to reduced density
matrices, or ``shadows'' in restricted subspaces. If we take a
arbitrary (nonstationary) state $|\phi_0 \kt$ and evolve it according
to $|\phi(T)\kt\,=\, {\cal U}^T |\phi_0 \kt $.  The reduced density
matrix at any time $T$ cannot be derived based solely on the reduced
density matrices of individual states as:
\beqa
\rho_1(T)\,=\, \mbox{tr}_2(|\phi(T)\kt \br \phi(T)|) \,&=& \, 
\sum_k |\br \psi_k|\phi_0 \kt |^2 \, \rho_{1k} \, +\,\nonumber \\ && 
\sum_{k \ne l}\exp(i(\psi_k-\psi_l)T)\br \psi_k|\phi_0\kt \br \phi_0|\psi_l\kt
\mbox{tr}_2(|\psi_k \kt \br \psi_l |),
\eeqa
where $\psi_k$ are eigenangles of the corresponding eigenstates.  If
we assume a non-degenerate spectrum, the reduced density matrix {\it
averaged over all time} is the first sum of the above expression and
is simply a weighted sum over the reduced density matrices of
individual eigenstates.

Thus we expect, based on our previous discussion of diagonally
dominant density matrices of eigenstates, that any arbitrary state's
reduced density matrix will rapidly evolve to a predominantly diagonal
one. This is of course reminiscent of ``decoherence'' phenomena
\cite{Joos,Zurek} and indeed as far as each degree of freedom individually
is concerned, it is an open system from which phase information can
flow out or decohere. However, the state will evolve in such a way
that even the diagonal part of the density matrix is significantly
altered, {\it i.e.,} it is not comparable to a situation wherein a
pure state is ``reduced'' to a classical ensemble by a measurement
like process.  Fig.~(\ref{figtime}) shows the evolution of an initial
state which is the sum of two well separated Gaussian in position
representation. The initial reduced density matrix has large
off-diagonal parts. The evolution under a weak coupling produces
interesting structures. After even one time step, for low couplings we
see the density matrix having the off-diagonal parts significantly
reduced. In fact this picture ``looks'' like the one resulting from
decoherence \cite{Zurek}.  However we are doing a numerically exact
computation and not using any approximate master equation. For later
times the case of small coupling or low chaos leads to fairly
non-diagonal density matrices, while in comparison the right panel
shows the evolution of the same initial state for the first four time
units and one sees that effective diagonality is rapidly achieved.

The time evolving density matrices represent the ``Schmidt paths''
\cite{Albrecht}.  If we start with an initially disentangled state,
the dynamics when completely chaotic, will quickly rotate the state
into those in which there is maximal entanglement in some sense. Is
this entanglement different from that observed for stationary states?
In fact from numerical calculations not shown here it seems that these
are identical, modulo fluctuations. This is unlike the case of the
Shannon entropy which for a time evolving pure state is different from
a stationary state, basically due to the fact that the time evolving
state is in general complex while (for time-reversal symmetric systems
as we are currently discussing) the stationary state is real. The
typical behaviour of the entropy is as expected a rise from zero to
the constant value of $\log(0.59 N)$ with a rapidity that is a
function of the coupling strengths and hence of the chaos in the
system.

We briefly also comment on the case when time reversal invariance is
broken.  That is achieved in the model studied in this paper by
choosing a non-zero $\beta$, which is equivalent to an introduction of
a magnetic flux line. While of course much of what has been said
already carries over to this case, there may be quantitative
differences. For one, the average entanglement of eigenstates is
slightly higher at about $\log(0.61 N)$. Thus time-reversal breaking
interactions may on the average produce more entanglement, and the
reduced density matrices are sensitive to time-reversal breaking.

\section{Discussion}

We have presented a variety of essentially numerical results concerning 
reduced density matrices of chaotic systems. We have seen that entanglement 
or the von Neumann entropy can be a good measure of localization and that the 
Schmidt states provide graded information about the nature of localization.
We have seen how chaos aids entanglement and it is a small step to extend
the universality observed from quantum chaos to reduced density matrices.
However we have not sufficiently explored the density matrices to be able
to comment on their randomness or type of universality we may expect.

There are various ways in which the analysis can be extended. One
important direction would be to have more than two coupled maps, or in
general many particle interacting systems.  Such higher dimensional
systems force us to face several problems. One is that Schmidt
decomposition is no longer possible, and entanglement becomes much
harder to quantify.  Ways to measure entanglement in multipartite
systems have been proposed recently that may prove to be useful. The
other is the increasing complexity of the classical system (if there
is one), and the exponentially growing numerical task. In perhaps a
related vein entanglement inherent in the ground state of a
anti-ferromagnetic model has been recently studied
\cite{Wootters}. Loss of quantum coherence was studied via simple models
in \cite{Albrecht} by coupling two state systems to larger
Hamiltonians with matrix representations having randomly chosen
elements. A natural situation in this context and in the spirit of
this paper is to look at bipartite quantum chaotic systems as we have
but whose dimensionality is unequal.
 
An important step toward a fundamental understanding of why random
matrix modeling must be successful at all was achieved by the use of
semiclassical methods and periodic orbits sums \cite{Berry}.  A
further direction would be to derive and use semiclassical orbit sums
for ``partial traces''. This obviously has close links with the
Feynman-Vernon path integral treatment of quantum dissipation. We note
that
\beq
\sum_{T}\exp(-i \psi T)\, \mbox{tr}_2 ({\cal U}^T)\, =\, \sum_k \sum_m \delta(
\psi \, -\, \psi_k \, -\, 2 \pi m )\, \rho_{1k},
\eeq
where $\psi_k$ is the eigenangle corresponding to the state whose
reduced density matrix is $\rho_{1k}$. The partial trace of the
propagator is therefore naturally a quantity of interest. The
classical orbits that will be the stationary paths will be
``partially-periodic'' orbits which for a given time $T$ connect
two configuration space points in subsystem $1$ while appearing
periodic in subsystem $2$.  Such semiclassical analysis of partial
traces may provide deeper understanding of various aspects of quantum
open systems in general.

\newpage

\begin{center}
{\bf \large FIGURES}
\end{center}

\begin{figure}
\caption{Average entropy as a function of the  coupling b. From top to bottom
the cases correspond to $N=25,\,20 \,\mbox{and} \,15$ respectively. The 
inset is a log-log plot of the same.}
\label{figavent}
\end{figure}

\begin{figure}
\caption{Entropy of all the 1600 states for the case $K_1=0.1$, $K_2=0.15$,
$b=2$ and $N=40$.The straight line is at the value $\log(0.59 N)$, while 
the states below an entropy of $3$ are indicated by circles. The localized 
states $(i=212)$ is the second encircled point.   } 
\label{figent}
\end{figure}

\begin{figure}
\caption{Eigenfunctions of a typical state (top, $i=6$) and a localized 
state (bottom, $i=212$). Parameter values are same as that of the previous 
figure} 
\label{figefn}
\end{figure}

\begin{figure}
\caption{The reduced density matrices, obtained by tracing out the second degree of freedom, 
for the delocalized 
 state (top, $i=6$) and the localized 
state (bottom, $i=212$) of the previous figure.}
\label{figrdm}
\end{figure}

\begin{figure}
\caption{The spectral average of the square of the elements of the 
 reduced density matrices  for
the cases of $b=0.05$ (above) and $b=2$ (below). $N=20$, and
$K_1=0.1$ and $K_2=0.15$ in this figure}
\label{figavRDM}
\end{figure}

\begin{figure}
\caption{Principal eigenvalues of the two reduced density matrices 
in Fig.~(\ref{figrdm}) are shown with the open circles and the 
closed circles corresponding to the delocalized and localized states 
respectively. Note that the scale is log-linear}
\label{figeval}
\end{figure}

\begin{figure}
\caption{The six principal Schmidt vectors for a typical chaotic state 
(left) and for the  localized state $(i=212)$ (right).}
\label{figvecrdm}
\end{figure}

\begin{figure}
\caption{The Husimi representation $W(q,p)$ of the principal Schmidt vectors
of the localized state ($i=212$). Top corresponds to tracing out the second
degree of freedom and the bottom the first.}
\label{fighusm}
\end{figure}

\begin{figure}
\caption{Time evolution of an initial reduced density matrix
shown in the left topmost figure. The left panels correspond to the
case $b=0.1$ for evolution over three time steps while the right 
corresponds to $b=2$ for evolution over four time steps.}
\label{figtime}
\end{figure}

\setcounter{figure}{0}

\newpage
\begin{figure}[h]
\hspace*{-2.5cm}\epsfbox{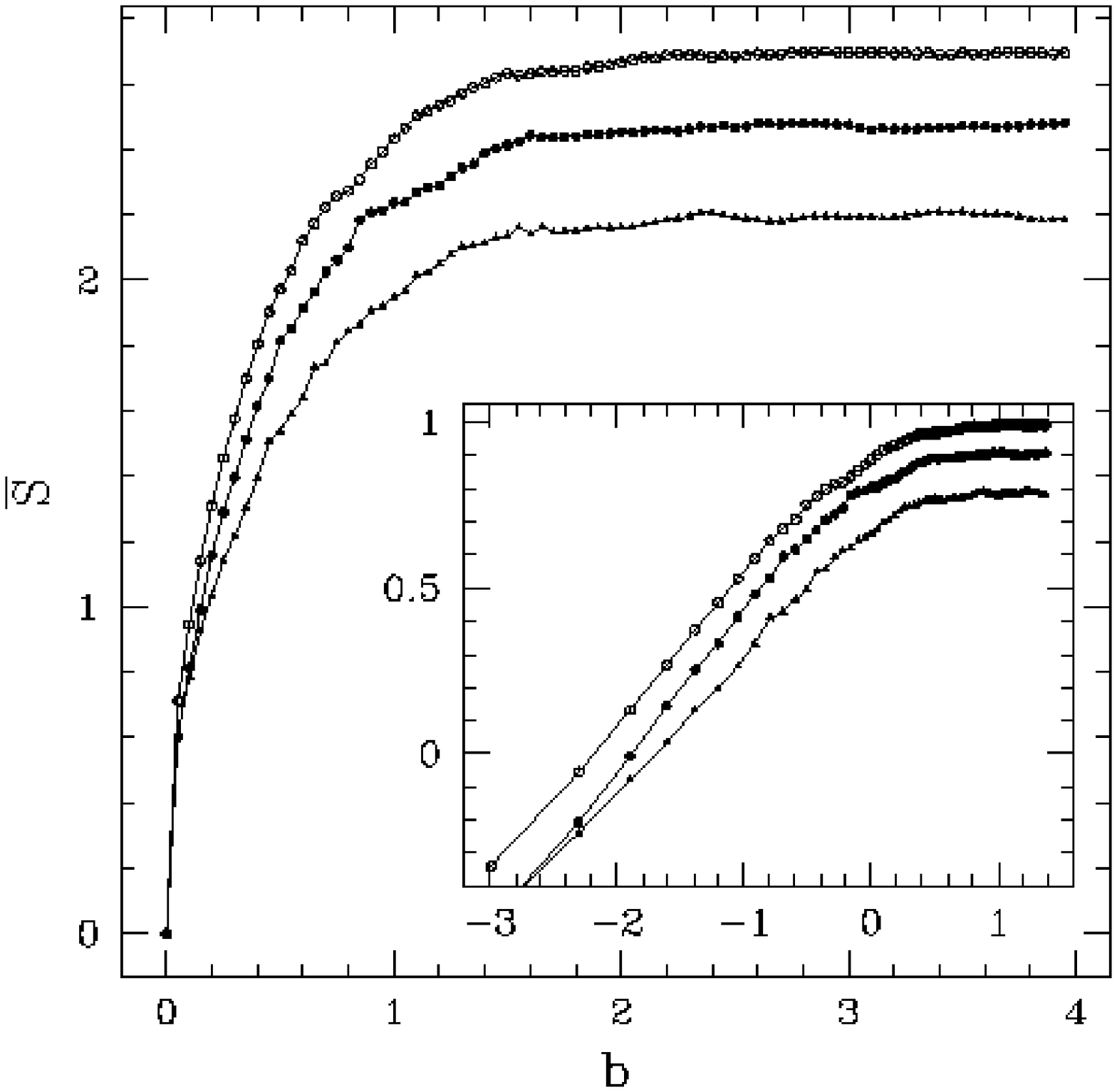}
\caption{}
\end{figure}

\newpage
\begin{figure}[h]
\hspace*{-2.5cm}\epsfbox{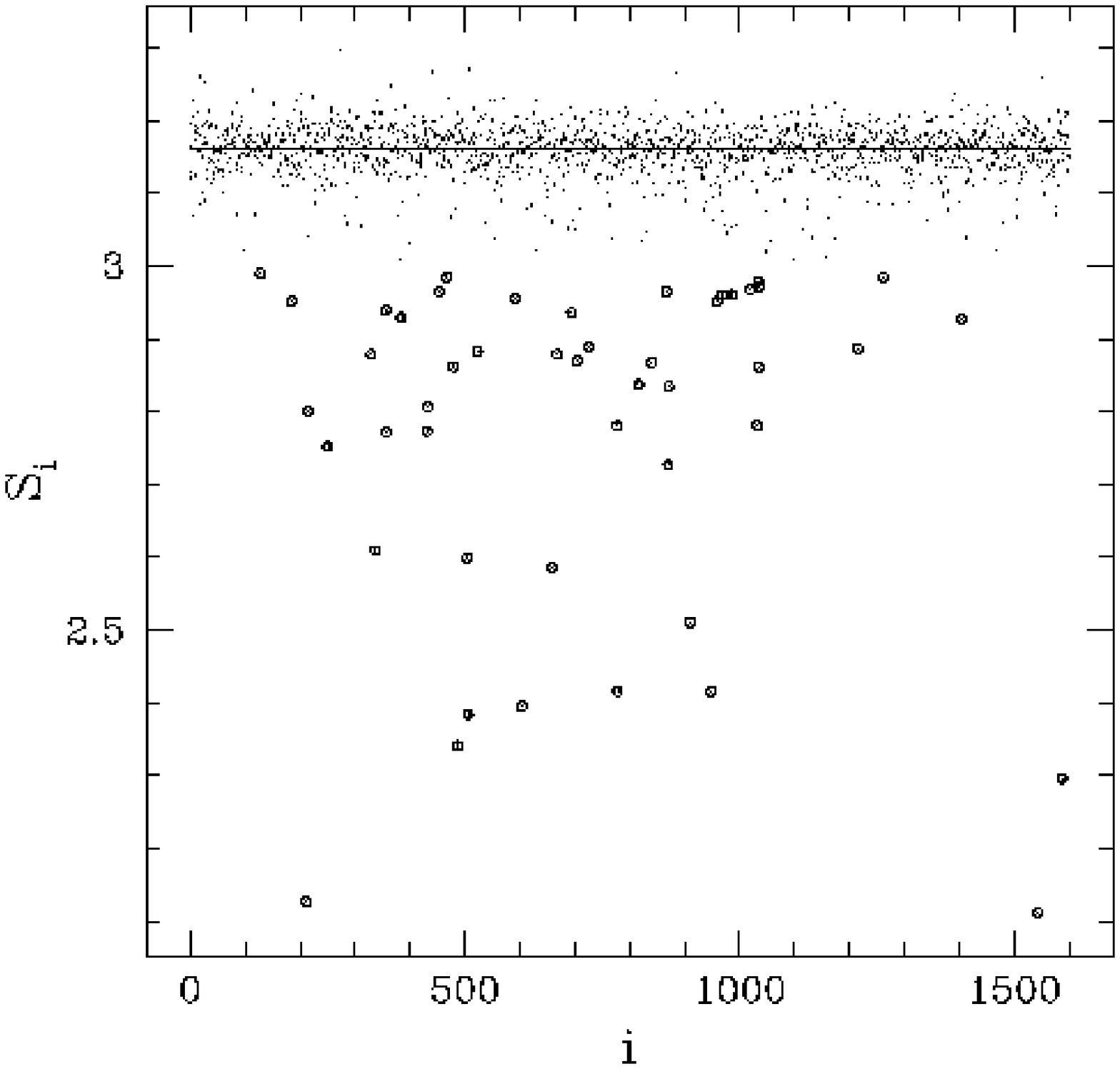}
\caption{}
\end{figure}

\newpage
\begin{figure}
\epsfbox{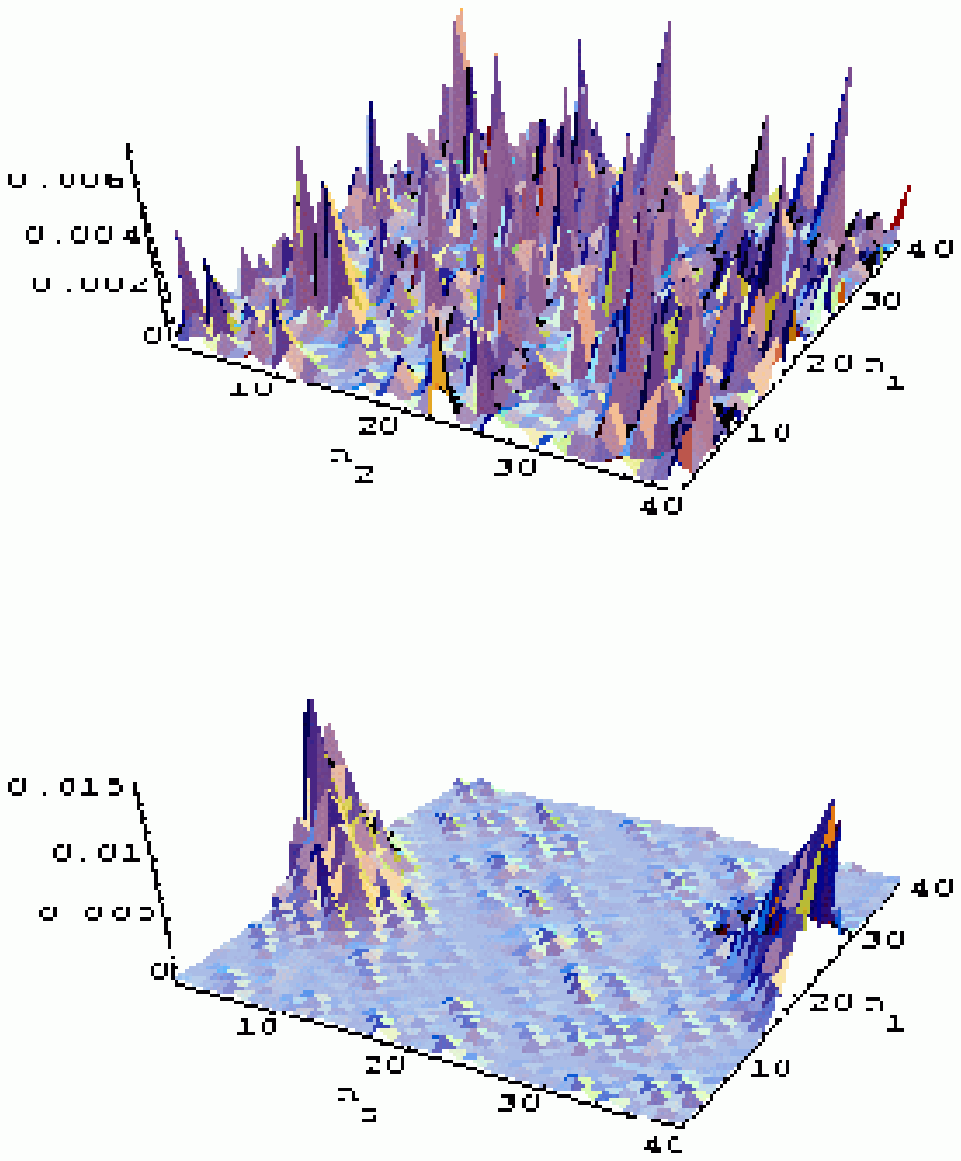}
\caption{}
\end{figure}

\newpage
\begin{figure}
\epsfbox{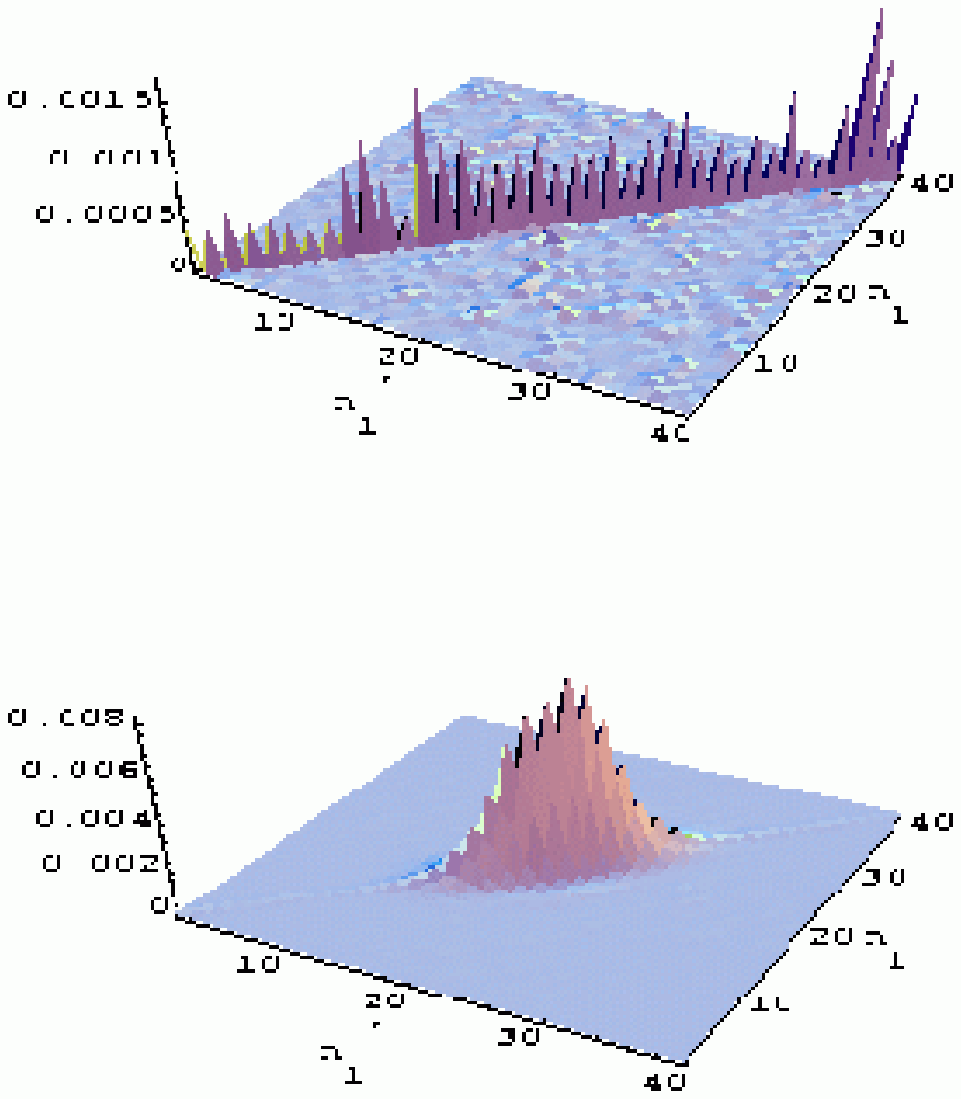}
\caption{}
\end{figure}

\newpage
\begin{figure}
\epsfbox{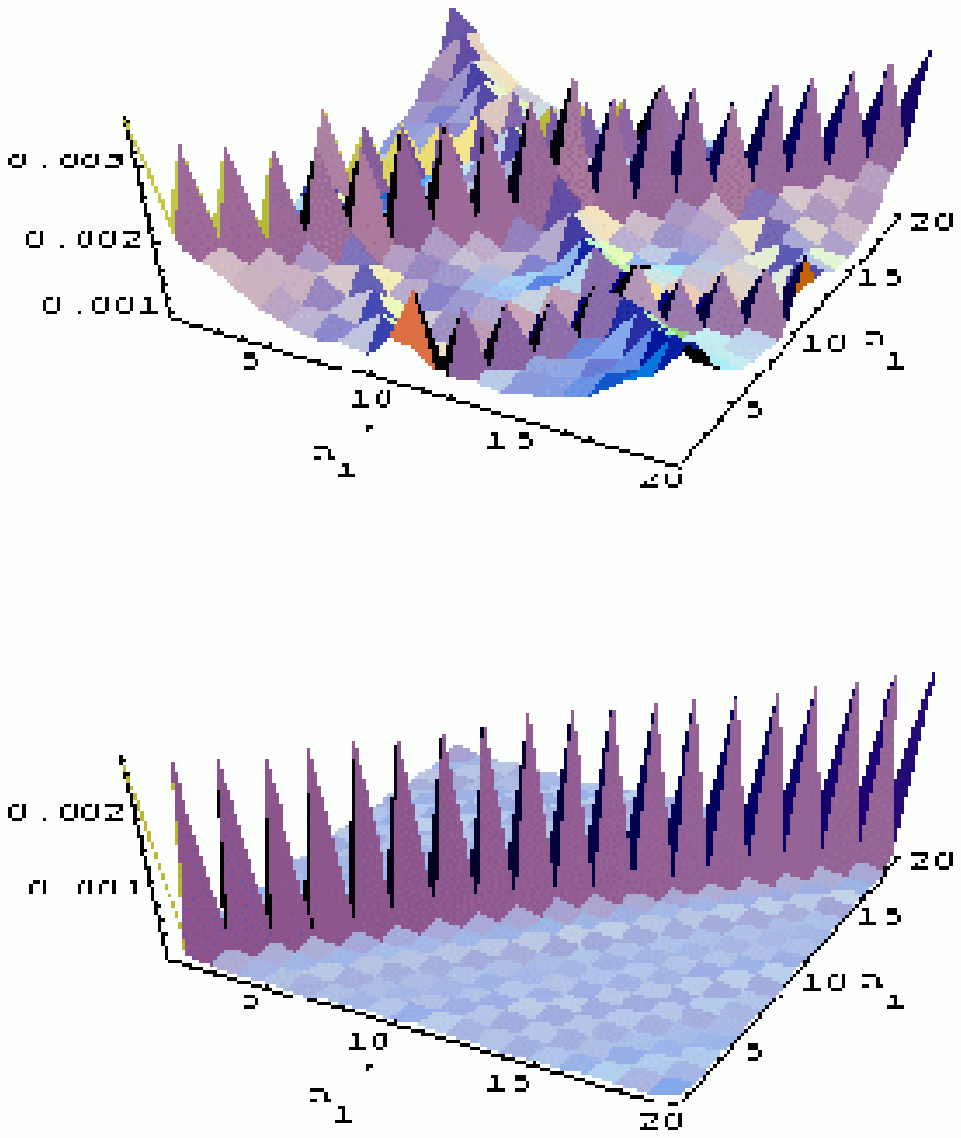}
\caption{}
\end{figure}

\newpage
\begin{figure}[h]
\hspace*{-2.5cm}\epsfbox{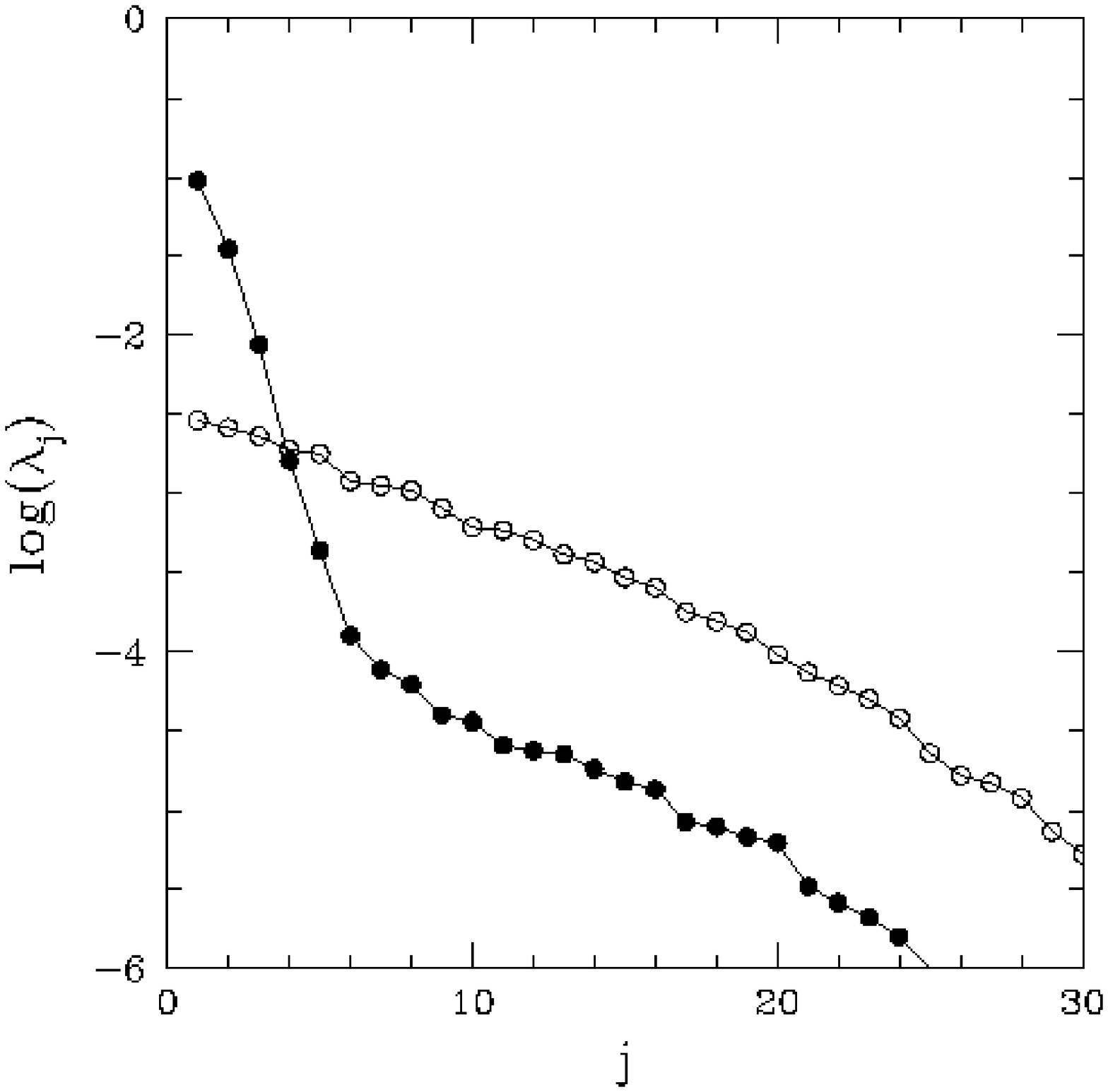}
\caption{}
\end{figure}

\newpage
\begin{figure}[h]
\hspace*{-2.5cm}\epsfbox{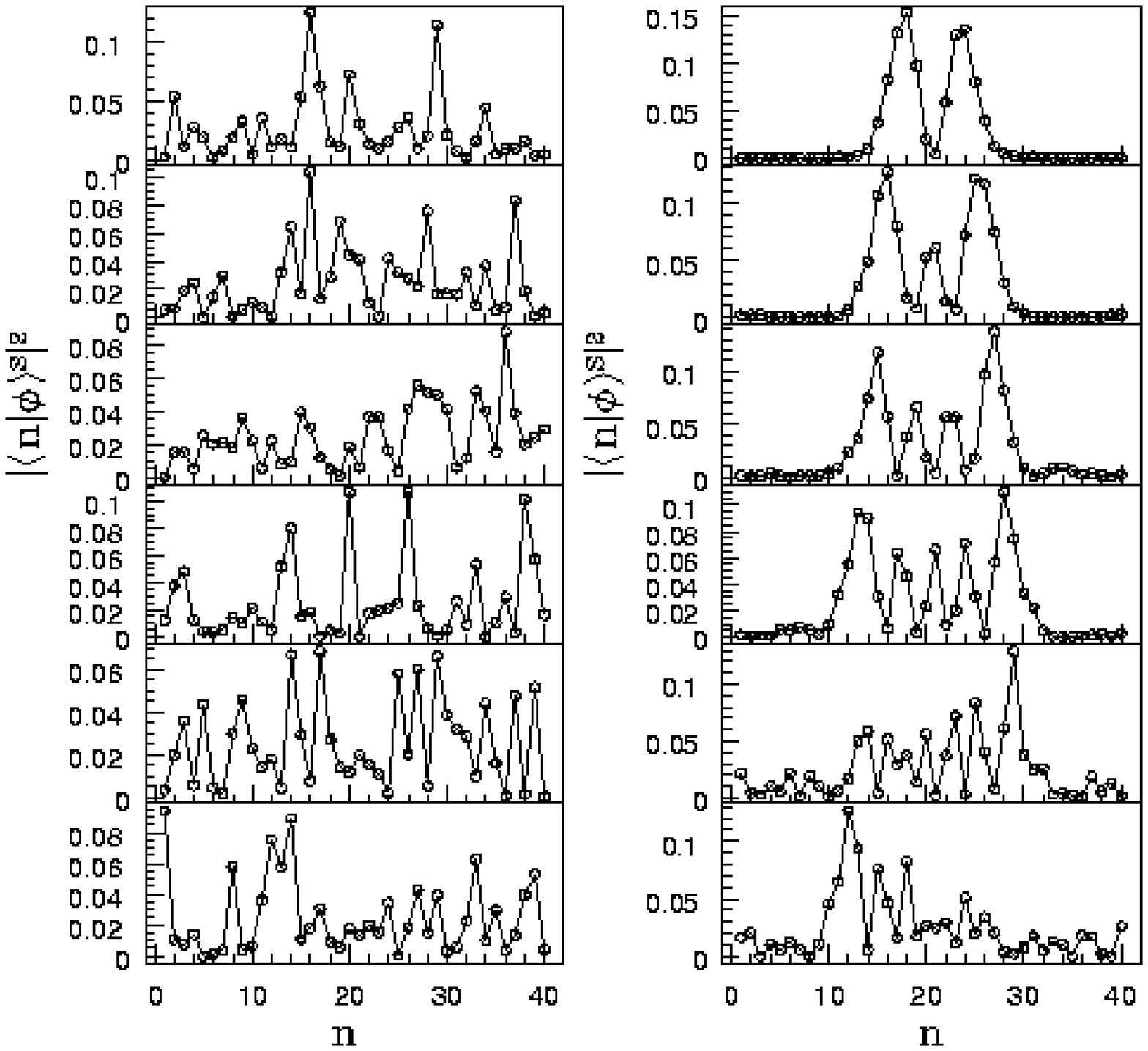}
\caption{}
\end{figure}

\newpage
\begin{figure}
\epsfbox{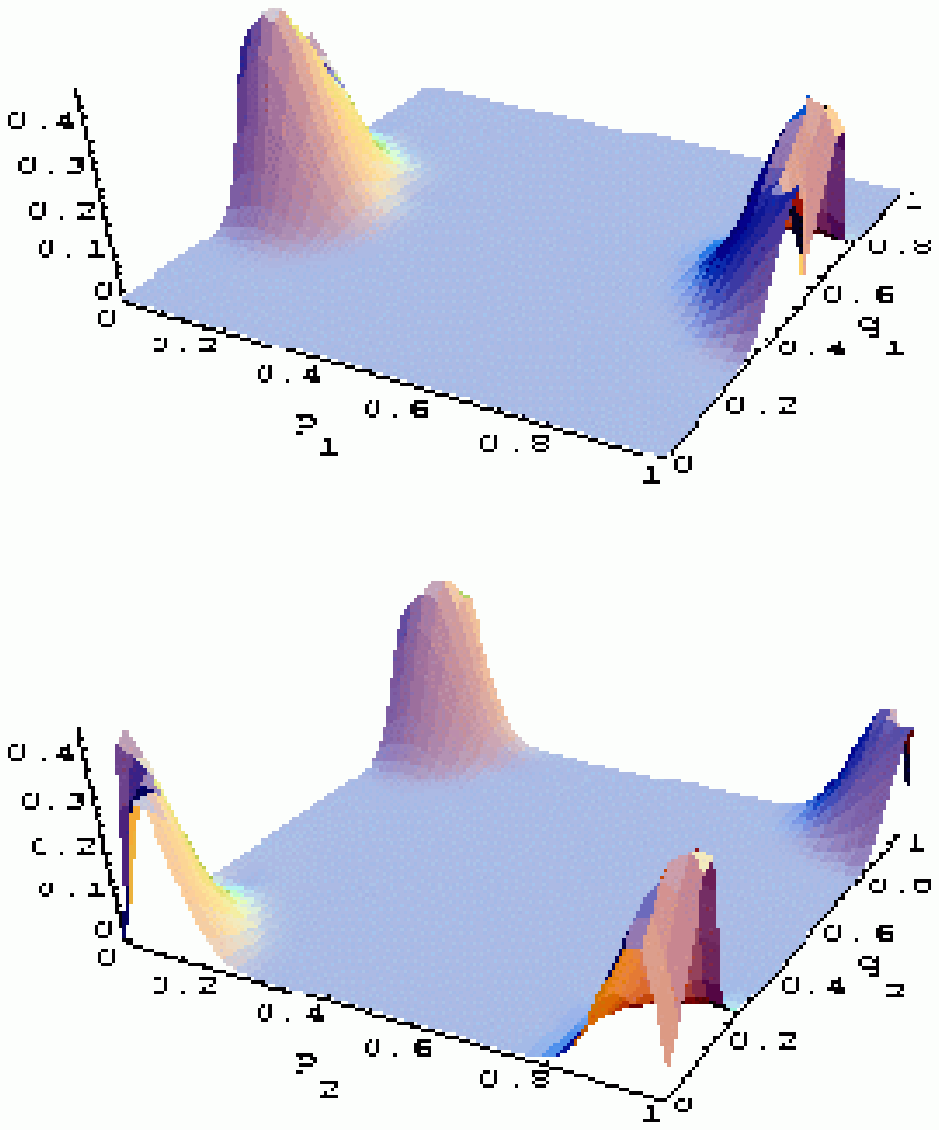}
\caption{}
\end{figure}

\newpage
\begin{figure}
\epsfbox{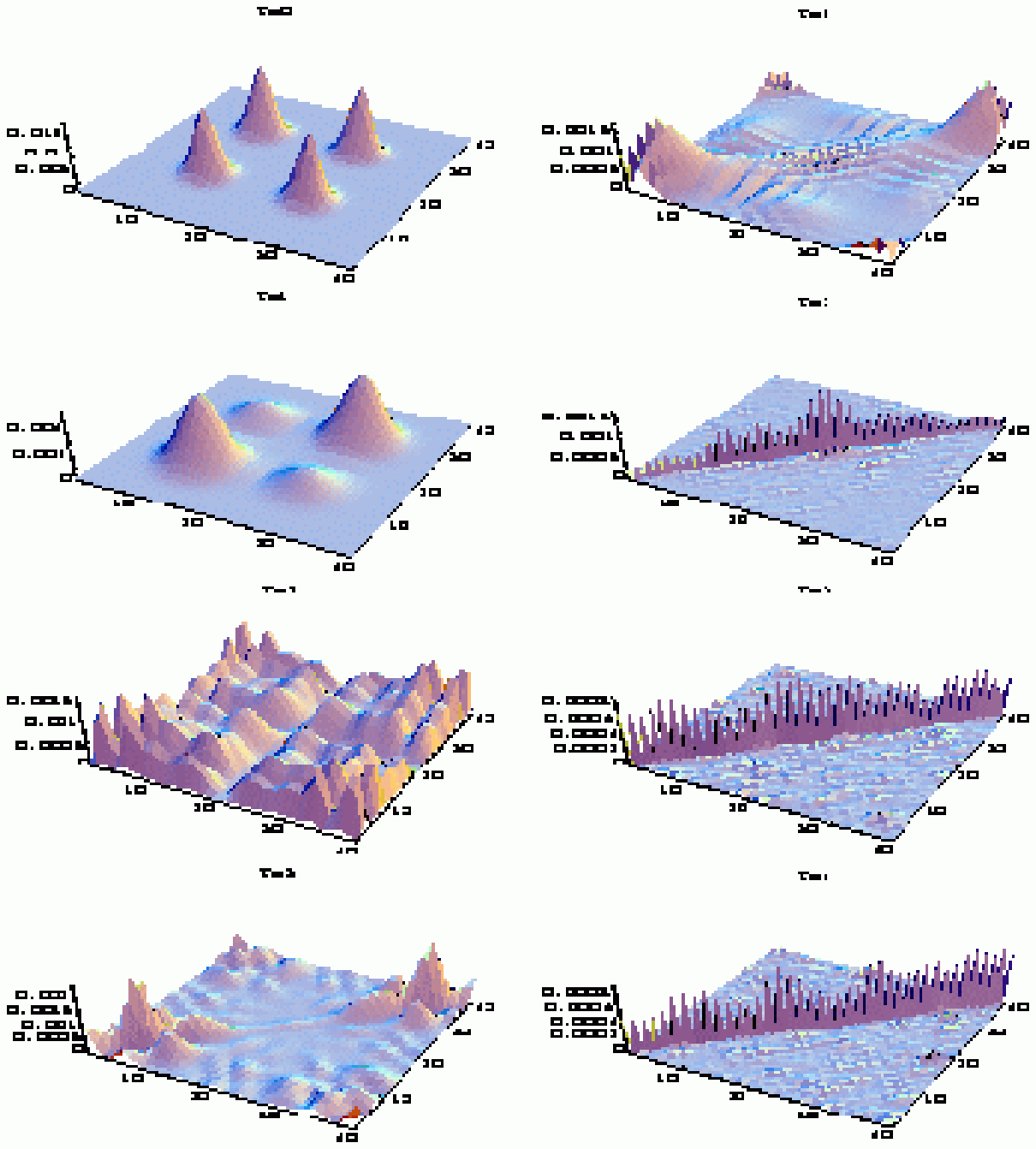}
\caption{}
\end{figure}


\begin{thebibliography}{99}

\bibitem{Peres} A.~Peres, {\it Quantum Theory: Concepts and Methods},
(Kluwer Academic Publishers, Dordrecht, 1993).

\bibitem{Steane} Reviewed by A.~ Steane, Rep. Prog. Phys. {\bf 61}, 117 (1998).

\bibitem{LL} A.~J.~Lichtenberg and M.~A.~Lieberman, {\it Regular and Chaotic Dynamics}, (Springer-Verlag, New York, 1992).

\bibitem{QCbooks} 
 M.~C.~Gutzwiller, {\it Chaos in Classical and Quantum Mechanics},
 Springer (New York, 1990); L.~E.~Reichl, {\it The transition to Chaos in Conservative Classical Systems: Quantum Manifestations}, (Springer-Verlag, New York, 1992).

\bibitem{Joos} D.~Giuilini, E.~Joos, C.~Kiefer, J.~Kupsch, I.~-O.~Stamatescu and H.~D.~Zeh, {\it Decoherence and the Appearance of a Classical World in Quantum Theory} (Springer-Verlag, Berlin,  1996)

\bibitem{HB}  J.~H.~Hannay and M.~V.~Berry, Physica D {\bf 1}, 267  (1980).

\bibitem{BV} N.~L.~Balazs and A.~Voros, Ann. Phys. (N. Y.) {\bf 190}, 1 (1989).

\bibitem{Izrailev}  F.~M.~Izrailev, Phys. Rep. {\bf 196}, 299 (1990).

\bibitem{Raizen} F.~L.~Moore, J.~C.~Robinson, C.~F.~Bharucha, P.~E.~Williams and M.~G.~Raizen, 
Phys. Rev. Lett. {\bf 73}, 2974 (1994).

\bibitem{4DClass} C.~Froeschl\'{e}, Astrophys. Space Sci. {\bf 14}, 110 (1971);
K.~Kaneko and R.~J.~Bagley, Phys. Lett A{\bf 110}, 435 (1985); B.~Wood, A.~J.~Lichtenberg and M.~A.~Lieberman, Phys. Rev. A {\bf 42}, 5885 (1990).

\bibitem{PopRoh97} S.~Popescu and D.~Rohrlich, Phys. Rev. A {\bf 56}, 
R3319 (1997).

\bibitem{Albrecht} A.~Albrecht, Phys. Rev. D {\bf 46}, 5504 (1992).

\bibitem{SSL} M.~S.~Santhanam, V.~B.~Sheorey, and A.~Lakshminarayan, Phys. Rev. E {\bf 57}, 345 (1998).

\bibitem{PandeyRMP} T.~A.~Brody, {\it et. al.} Rev. Mod. Phys. {\bf 53}, 
385 (1981).

\bibitem{Heller} E.~J.~Heller, Phys. Rev. Lett. {\bf 53} (1994); 
and in {\it Quantum Chaos and Statistical Nuclear Physics} Ed. T.~H.~Seligman and H.~Nishioka (Springer-Verlag, Berlin, 1986).

\bibitem{Kaplan}  L.~Kaplan and E.~J.~Heller,  Ann. Phys. {\bf 264}, 
171 (1998).

\bibitem{LCS}  A.~Lakshminarayan, N.~R.~Cerruti and S.~Tomsovic,
Phys. Rev. E {\bf 60}, 3992 (1999).

\bibitem{Saraceno} M.~Saraceno, Ann. Phys. (N.Y.) {\bf 199}, 37, (1990).

\bibitem{Santh} G.~G.~de~Polavieja, F.~Borondo 
and R.~M.~ Benito, Int. J. Quant. Chem {\bf 51}, 555 (1994);
 M~.S.~Santhanam, Ph. D. thesis (Physical 
Research Laboratory, Ahmedabad, 1997).

\bibitem{LebVoros} P.~Lebouef and A.~Voros, J. Phys. A {\bf 23}, 1765 (1990).

\bibitem{Zurek} W.~H.~Zurek, Physics Today {\bf 44} (Oct.), 36 (1991).

\bibitem{Wootters} K.~M.~O'~Connor and W.~K.~Wootters, quant-ph/0009041.

\bibitem{Berry} M.~V.~Berry, Proc. Roy. Soc. Lond. A {\bf 400}, 229 (1985).

\end{thebibliography}
\end{document}